\documentclass[aps,preprintnumbers,showpacs,nofootinbib,superscriptaddress,floatfix,notitlepage]{revtex4-1}

\pdfoutput=1 

\usepackage{graphicx}
\usepackage{graphics}
\usepackage{braket}
\usepackage{bbold}
\usepackage{amssymb}
\usepackage{amsmath}
\usepackage{dcolumn}
\usepackage{bm}
\usepackage{slashed}
\usepackage{datetime}
\usepackage{mciteplus}
\usepackage{multirow}
\usepackage{color}
\usepackage{float}
\usepackage[utf8]{inputenc}
\usepackage[normalem]{ulem}
\usepackage{hyperref}
\hypersetup{
 linktocpage = true,
 linkcolor = blue,
 urlcolor = blue,
 colorlinks = true,
 linkcolor = blue,
 anchorcolor = blue,
 citecolor = blue,
 urlcolor = blue,
 pdfstartview = {XYZ null null 1.25} 
           }

\usepackage[normalem]{ulem} 

\newcommand{\tr}{\operatorname*{Tr}\nolimits}

\begin{document}
\allowdisplaybreaks[2]

\title{Transverse-momentum-dependent gluon distribution functions in a spectator model}

\author{Alessandro Bacchetta}
\email{alessandro.bacchetta@unipv.it}
\affiliation{Dipartimento di Fisica, Universit\`a di Pavia, via Bassi
  6, I-27100 Pavia}
\affiliation{INFN Sezione di Pavia, via Bassi 6,
  I-27100 Pavia, Italy}

\author{Francesco Giovanni Celiberto}
\email{francescogiovanni.celiberto@unipv.it}
\affiliation{Dipartimento di Fisica, Universit\`a di Pavia, via Bassi
  6, I-27100 Pavia}
\affiliation{INFN Sezione di Pavia, via Bassi 6,
  I-27100 Pavia, Italy}

\author{Marco Radici}
\email{marco.radici@pv.infn.it}
\affiliation{INFN Sezione di Pavia, via Bassi 6, I-27100 Pavia, Italy}

\author{Pieter Taels}
\email{pieter.taels@polytechnique.edu}
\affiliation{Centre de Physique Th\'eorique, \'Ecole Polytechnique, CNRS, I.P. Paris, F-91128 Palaiseau, France}

\begin{abstract}
We present a model calculation of transverse-momentum-dependent distributions (TMDs) of gluons in the nucleon. The model is based on the
assumption that a nucleon can emit a gluon, 
and what remains after the emission is treated as a single spectator particle. This spectator particle is considered to be on-shell, but its mass is allowed to take a continuous range of values, described by a spectral function. The nucleon-gluon-spectator coupling is described by an effective vertex containing two form factors. We fix the model parameters to obtain the best agreement with collinear gluon distributions extracted from global fits. We study the tomography in momentum space of gluons inside nucleons for various combinations of their polarizations. These can be used to make predictions of observables relevant for gluon TMD studies at current and future collider facilities. 
\end{abstract}

\date{\today, \currenttime}


\maketitle

\section{Introduction}
\label{s:intro}

Transverse-Momentum-dependent parton Distributions (TMDs) have been a subject of intense study in the last years (see Ref.~\cite{Angeles-Martinez:2015sea} for a recent review). Whereas several results have been obtained concerning quark TMDs, much less is known about gluons.

Gluon TMDs have been classified for the first time in Ref.~\cite{Mulders:2000sh} and later also in Refs.~\cite{Meissner:2007rx,Lorce:2013pza,Boer:2016xqr}. Their factorization, evolution and universality properties have been investigated in Refs.~\cite{Ji:2005nu,Buffing:2013kca,Boer:2014tka,Echevarria:2015uaa,Echevarria:2016scs}. 
Possible ways to access gluon TMDs in experiments have been proposed in the literature~\cite{Boer:2010zf,Pisano:2013cya,Boer:2016fqd,Zheng:2018awe,Sun:2011iw,Boer:2011kf,Yuan:2008vn,Godbole:2014tha,Mukherjee:2016qxa,Bacchetta:2018ivt,DAlesio:2019qpk,Dunnen:2014eta,Lansberg:2017tlc,Scarpa:2019fol,Lansberg:2017dzg}. 
Recent discussions on TMD factorization in quarkonium production have been presented in Refs.~\cite{Echevarria:2019ynx,Fleming:2019pzj,Boer:2020bbd}.

At low $x$, the so-called \emph{unintegrated gluon distribution} (UGD) has been the subject of intense investigations since the early days. A first definition for the UGD was given in the Balitsky--Fadin--Kuraev--Lipatov (BFKL) approach~\cite{Fadin:1975cb,Kuraev:1976ge,Kuraev:1977fs,Balitsky:1978ic}. 
Its precise relation to the small-$x$ limit of the unpolarized gluon TMD was established only recently~\cite{Dominguez:2010xd,Dominguez:2011wm}. 
An overview of the available literature on unpolarized and helicity gluon TMDs at low $x$ can be found in Ref.~\cite{Petreska:2018cbf} (and references therein). Some very recent theoretical and phenomenological studies are discussed in Refs.~\cite{Altinoluk:2018byz,Altinoluk:2020qet,Yao:2018vcg,Zhou:2018lfq,Altinoluk:2019wyu}. 

Accessing gluon TMDs is one of the primary goals of new experimental facilities~\cite{Boer:2011fh,Accardi:2012qut,Brodsky:2012vg,Aidala:2019pit}. 
In this exploratory context, it is particularly useful to develop models for gluon TMDs. Models can be employed, for example, to expose qualitative features of
gluon TMDs, confirm or falsify generally accepted assumptions, make reasonable predictions for experimental observables, or guide the choice of
functional forms to be used in gluon TMD fits. 

Quark TMD models have been widely used for these purposes in the past (see, \emph{e.g.},
Refs.~\cite{Jakob:1997wg,Brodsky:2002cx,Gamberg:2005ip,Gamberg:2007wm,Goeke:2006ef,Meissner:2007rx,Bacchetta:2008af,Pasquini:2008ax,Bacchetta:2010si,Avakian:2010br,Lorce:2011zta,Burkardt:2015qoa,Kovchegov:2015zha,Pasquini:2019evu}). Effective models of the UGD can be found in Refs.~\cite{GolecBiernat:1998js,Ivanov:2000cm,Kimber:2001sc,Hentschinski:2012kr,Kutak:2012rf,Hautmann:2013tba}.
Predictions based on some of these models have been compared to experimental data for the exclusive diffractive vector-meson leptoproduction at {\tt HERA}~\cite{Anikin:2011sa,Besse:2013muy,Bolognino:2018rhb,Bolognino:2019pba,Celiberto:2019slj} and for the inclusive forward Drell--Yan dilepton production at {\tt LHCb}~\cite{Brzeminski:2016lwh,Motyka:2016lta,Celiberto:2018muu}. 
Conversely, very little has been done for gluon TMDs at intermediate $x$. Model calculations of these functions have been discussed only in Refs.~\cite{Pereira-Resina-Rodrigues:2001eda,Meissner:2007rx,Lu:2016vqu}. 

In this work, we present an extension of our spectator-model calculation of quark TMDs~\cite{Bacchetta:2008af} to unpolarized and polarized ($T$-even)
gluon TMDs, effectively incorporating also small-$x$ effects. The model is based on the assumption that a nucleon can emit a gluon, 
after which the remainders are treated as a single spectator particle. The nucleon-gluon-spectator coupling is described by an effective vertex containing two form factors. At variance with our previous work, the spectator mass can take a continuous range of values described by a spectral function. We determine the parameters of the model by reproducing the gluon unpolarized and helicity collinear parton distribution functions (PDFs) obtained in global fits.  

The paper is structured as follows. In Sec.~\ref{s:model}, we highlight the main features of our spectator model. In Sec.~\ref{s:params}, we describe how we fix the model parameters by getting the best possible agreement with collinear gluon PDFs obtained in global fits. In Sec.~\ref{s:out}, we show our model results for all the $T$-even gluon TMDs. Finally, in Sec.~\ref{s:end} we draw our conclusions and discuss some outlooks.

\section{Formalism}
\label{s:model}

We represent a generic 4-momentum $a$ through its light-cone components $[ a_-, \, a_+, \bm{a}_T ]$, where $a_\pm = a \cdot n_\mp$ and $n_\pm$ are light-like vectors satisfying $n_\pm^2 = 0$ and $n_+ \cdot n_- = 1$. 
Following Ref.~\cite{Bacchetta:2008af}, we work in the frame where the nucleon
momentum $P$ has no transverse component, i.e., 
\begin{equation}
P = \left[ \frac{M^2}{2 P^+}, \, P^+ , \, \bm{0} \right] \; ,
\label{eq:frame}
\end{equation}
where $M$ is the nucleon mass. 
The parton momentum is parametrized as 
\begin{equation}
p = \left[ \frac{p^2 + \bm{p}_T^2}{2 x P^+} , \, x P^+ , \, \bm{p}_T \right] \; ,
\label{eq:parton}
\end{equation}
where evidently $x = p^+ / P^+$ is the light-cone (longitudinal) momentum fraction carried by the parton. For the nucleon state $|P, S \rangle$ with momentum $P$ and spin $S$, the gauge-invariant gluon-gluon correlator reads~\cite{Mulders:2000sh} 
\begin{equation}
\Phi^{\mu \nu, \rho \sigma} (x, \bm{p}_T; S) = \frac{1}{x P^+} \int \frac{d\xi^- d\bm{\xi}_T}{(2 \pi )^3} \, e^{i p\cdot \xi} \, \langle P, S | F^{\rho \sigma}_a (0) \, \mathcal{U}_{ab} (0, \xi ) \, F^{\mu \nu}_b (\xi) | P, S \rangle \vert_{\xi^+ = 0} \; , 
\label{eq:correlator}
\end{equation}
where (here, and in the following) a summation upon repeated (color) indices
is understood.
The field tensor $F_a^{\mu \nu}$ is related to the gluon field $A_a^\mu$ by
$F_a^{\mu \nu} = \partial^\mu A_a^\nu - \partial^\nu A_a^\mu + g \, f_{abc} \,
A_b^\mu \, A_c^\nu$, with $f_{abc}$ the structure constants of the color SU(3)
group and $g$ the strong coupling.
The symbol $\mathcal{U}_{ab} (0, \xi )$ denotes the gauge-link operator
\begin{equation}
\mathcal{U}_{ab} (0, \xi )  = \mathcal{P} \, \exp \left[ - g \, f_{abc} \int_0^\xi dw \cdot A_c (w) \right] \; , 
\label{eq:gaugelink}
\end{equation}
which connects the two different space-time points $0$ and $\xi$
along a path that is determined by the process.
There are at least two possible definitions of the correlator that involve two
different gauge-link choices, leading to the so-called Weizs\"acker-Williams
(WW) and 
dipole gluon TMDs~\cite{Kharzeev:2003wz,Dominguez:2011wm}, which can be probed
in different processes.     
In this work, we consider only
leading-order contributions, neglecting the effect of the gauge link and
its process dependence (see, e.g., Ref.~\cite{Boer:2016xqr,Boer:2016bfj}).\footnote{We recall that in the Weizs\"acker--Williams
  representation it is always possible to choose a gauge where the gauge-link
  operator reduces to unity~\cite{Dominguez:2010xd,Dominguez:2011wm}}
Therefore, our calculation at the present stage of sophistication can be
considered to be a model for both definitions of gluon TMDs.

\begin{figure}[h]
 \includegraphics[width=0.5\textwidth]{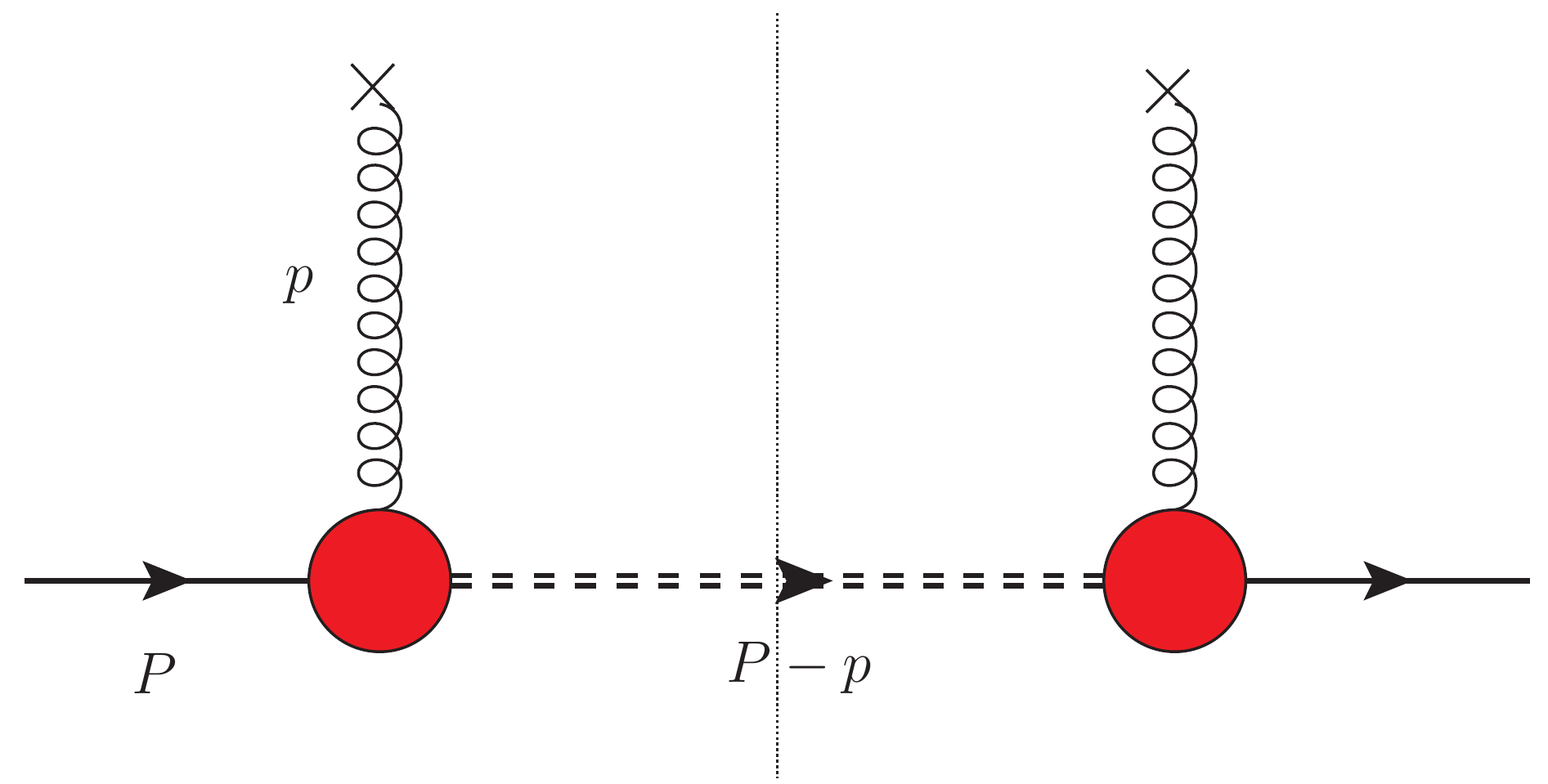}
 \caption{Tree-level cut diagram for the calculation of $T$-even leading-twist
   gluon densities. The double dashed line represents the
   spin-$\textstyle{\frac{1}{2}}$ spectator. The red blob represents the
   nucleon-gluon-spectator vertex. Gluon lines with crosses correspond to
   specific Feynman rules for the gluon field tensor (see text).}
\label{fig:diagram}
\end{figure}

In the following, we consider the leading-twist component of the gluon-gluon correlator $\Phi^{+ i, + j} \equiv \Phi^{i j}$ with $i, j$ transverse spatial indices~\cite{Mulders:2000sh}. We evaluate it in the spectator approximation, namely we assume that the nucleon in the state $|P, S \rangle$ can split into a gluon with momentum $p$ and other remainders, effectively treated as a single spin-$\textstyle{\frac{1}{2}}$ spectator particle with momentum $P-p$ and mass $M_X$. Similarly to Ref.~\cite{Bacchetta:2008af}, we define a ``tree-level" scattering amplitude $\mathcal{M}_a^i (S)$ given by (see Fig.~\ref{fig:diagram}) 
\begin{eqnarray}
\mathcal{M}_a^i (S) &= &\langle P-p | F_a^{+ i} | P, S \rangle \nonumber \\
&= &\bar{u}_c (P-p) \, \frac{G_{a b}^{i \mu} (p,p)}{p^2} \, \mathcal{Y}_{\mu, b c} \, U (P,S) \; , 
\label{eq:scattampl}
\end{eqnarray}
where $U$ is the nucleon spinor and $\bar{u}_c$ is the spinor of the spectator
with color $c$. The term
\begin{equation} 
  G_{a b}^{i \mu} (p,k)
  = - i \delta_{ab}  \biggl(g^{i \mu} -\frac{ k^i (n_-)^{\mu}}{p^+}\biggr)
\end{equation}
represents a specific Feynman rule
for the field tensor in the definition of the
correlator~\cite{Goeke:2006ef,Collins:2011zzd}. We remark that at the
accuracy we are working all results are gauge invariant. 

We model the nucleon-gluon-spectator vertex as
\begin{equation}
\mathcal{Y}_{b c}^\mu = \delta_{b c} \, \bigg[ g_1 (p^2) \, \gamma^\mu + g_2 (p^2) \, \frac{i}{2 M}\, \sigma^{\mu \nu} \, p_\nu \bigg] \; , 
\label{eq:vertex}
\end{equation}
where as usual $\sigma^{\mu \nu} = i [ \gamma^\mu , \gamma^\nu ] /2$, and $g_{1,2} (p^2)$ are model-dependent form factors. 
With our assumptions the spectator is identified with an on-shell spin-$\textstyle{\frac{1}{2}}$ particle, much like the nucleon. Although in principle the expression of $\mathcal{Y}^\mu$ could contain more Dirac structure, we model it similarly to the conserved electromagnetic current of a free nucleon obtained from the Gordon decomposition. 
The form factors $g_{1,2} (p^2)$ are formally similar to the
Dirac and Pauli form factors, but obviously must not be identified with them. Consistently with our previous model description of quark TMDs~\cite{Bacchetta:2008af}, we use the dipolar expression 
\begin{equation}
g_{1,2} (p^2) = \kappa_{1,2} \, \frac{p^2}{|p^2 - \Lambda_X^2|^2} = \kappa_{1,2} \, \frac{p^2 \, (1 - x)^2}{(\bm{p}_T^2 + L_X^2 (\Lambda_X^2))^2}  \; ,
\label{eq:dipolarff}
\end{equation}
where $\kappa_{1,2}$ and $\Lambda_X$ are normalization and cut-off parameters, respectively, and 
\begin{equation}
L_X^2 (\Lambda_X^2) = x \, M_X^2 + (1 - x) \, \Lambda_X^2 - x \, (1 - x) \, M^2 \; .
\label{eq:LX}
\end{equation}
The dipolar expression of Eq.~\eqref{eq:dipolarff} has several advantages: it cancels the singularity of the gluon propagator, it smoothly suppresses the effect of high $\bm{p}_T^2$ where the TMD formalism cannot be applied, and it compensates also the logarithmic divergences arising after integration upon $\bm{p}_T$. 

Using Eq.~\eqref{eq:scattampl}, we can write our spectator model approximation to the gluon-gluon correlator at tree level as
\begin{eqnarray}
\Phi^{i j} (x, \bm{p}_T, S) &\sim &\frac{1}{(2 \pi)^3} \, \frac{1}{2\, (1-x)\, P^+} \, \tr \left[ \bar{\mathcal{M}}_a^j (S) \, \mathcal{M}_a^i (S) \right] \, \vert_{p^2 = \tau (x, \bm{p}_T^2)} \nonumber \\
&= &\frac{1}{(2 \pi)^3} \, \frac{1}{2\, (1-x)\, P^+} \, \tr \left[ (\slashed{P} + M) \,  \frac{1 + \gamma^5 \slashed{S}}{2} \, G_{a b'}^{j \nu *} (p,p) \, \mathcal{Y}_{\nu, b' c'}^* \, G_{a b}^{i \mu} (p,p) \, \mathcal{Y}_{\mu, b c} \, (\slashed{P} - \slashed{p} + M_X)_{c c'} \right] \; ,
\label{eq:correlator_spect} 
\end{eqnarray}
where a trace upon color and spinorial indices is understood. The assumed on-shell condition $(P-p)^2 = M_X^2$ for the spectator implies that the gluon is off-shell by
\begin{equation}
p^2 \equiv \tau (x, \bm{p}_T^2) = - \frac{\bm{p}_T^2 + L_X^2 (0)}{1-x} \; .
\label{eq:offshell}
\end{equation}

The leading-twist $T$-even gluon TMDs can be obtained by suitably projecting $\Phi^{i j}$~\cite{Mulders:2000sh,Meissner:2007rx}:
\begin{eqnarray}
\hat{f}_1^g (x, \bm{p}_T^2; M_X) &= &- \frac{1}{2}\, g^{i j} \, \left[ \Phi^{i j} (x, \bm{p}_T, S) + \Phi^{i j} (x, \bm{p}_T, -S) \right]  \nonumber \\
&= &\Big[ \big( 2 M x g_1 - x (M+M_X) g_2 \big)^2 \, \big[ (M_X - M (1-x) )^2 + \bm{p}_T^2 \big] + 2 \bm{p}_T^2 \, (\bm{p}_T^2 + x M_X^2) \, g_2^2 \nonumber \\
& &\quad + 2 \bm{p}_T^2 M^2 \, (1-x) \, (4 g_1^2 - x g_2^2 ) \Big]  \nonumber \\ 
& &\times \Big[ (2 \pi)^3 \, 4 x M^2 \, (L_X^2 (0) + \bm{p}_T^2)^2 \Big]^{-1} \; ,
\label{eq:f1}
\end{eqnarray}

\begin{eqnarray}
\hat{g}_{1L}^g (x, \bm{p}_T^2; M_X) &= &\frac{1}{S_L} \, i \varepsilon_T^{i j} \, \Phi^{i j} (x, \bm{p}_T, S_L)  \nonumber \\
&= &\Big[ \bm{p}_T^2 \, (2-x)\, \big( 2 M \, g_1 - (M_X - M) \, g_2 \big) + 2 M x \, \large( M_X - M\, (1-x) \large)^2 \, g_1 \nonumber \\
& &\quad - \big[ (M_X+M)\, x\, \big( L_X^2 (0) + (1-x)\, (M_X-M)^2 \big) + 2 M x \, \bm{p}_T^2 \big] \, g_2 \Big]  \, \Big( 2 M \, g_1 - (M_X + M)\, g_2 \Big) \nonumber \\
& &\times \Big[ (2 \pi)^3 \, 4 M^2 \, (L_X^2 (0) + \bm{p}_T^2)^2 \Big]^{-1} \; ,
\label{eq:g1L}
\end{eqnarray}

\begin{eqnarray}
\hat{g}_{1T}^g (x, \bm{p}_T^2; M_X) &= &\frac{- M}{\bm{p}_T \cdot \bm{S}_T} \, i \varepsilon_T^{i j} \, \Phi^{i j} (x, \bm{p}_T, S_T)  \nonumber \\
&= &- \Big( 2 M \, g_1 - (M_X+M)\, g_2 \Big) \, \Big[ ( M_X - M\, (1-x) ) \, \big( 2 M\, (1-x) \, g_1 + x M_X \, g_2 \big) + \bm{p}_T^2 \, g_2 \Big] \nonumber \\
& &\times \Big[ (2 \pi)^3 \, 2 M \, (L_X^2 (0) + \bm{p}_T^2)^2 \Big]^{-1} \; ,
\label{eq:g1T}
\end{eqnarray}

\begin{eqnarray}
\hat{h}_1^{\perp g} (x, \bm{p}_T^2; M_X) &= &\frac{M^2}{\varepsilon_T^{i j} \delta^{j m} (p_T^j p_T^m + g^{j m} \bm{p}_T^2)} \, \varepsilon_T^{l n} \delta^{n r} \, \left[ \Phi^{n r} (x, \bm{p}_T, S) + \Phi^{n r} (x, \bm{p}_T, -S) \right] \nonumber \\
&= &\Big[ 4 M^2 \, (1-x) \, g_1^2 + ( L_X^2 (0) + \bm{p}_T^2 ) \, g_2^2 \Big] \nonumber \\
& &\times \Big[ (2 \pi)^3 \, x \, (L_X^2 (0) + \bm{p}_T^2)^2 \Big]^{-1} \; ,
\label{eq:h1perp}
\end{eqnarray}
where $\varepsilon_T^{i j} = \varepsilon^{- + i j}$ and $S_{L(T)}$ is the longitudinal (transverse) polarization of the nucleon.

\begin{figure}[h]
 \includegraphics[width=0.5\textwidth]{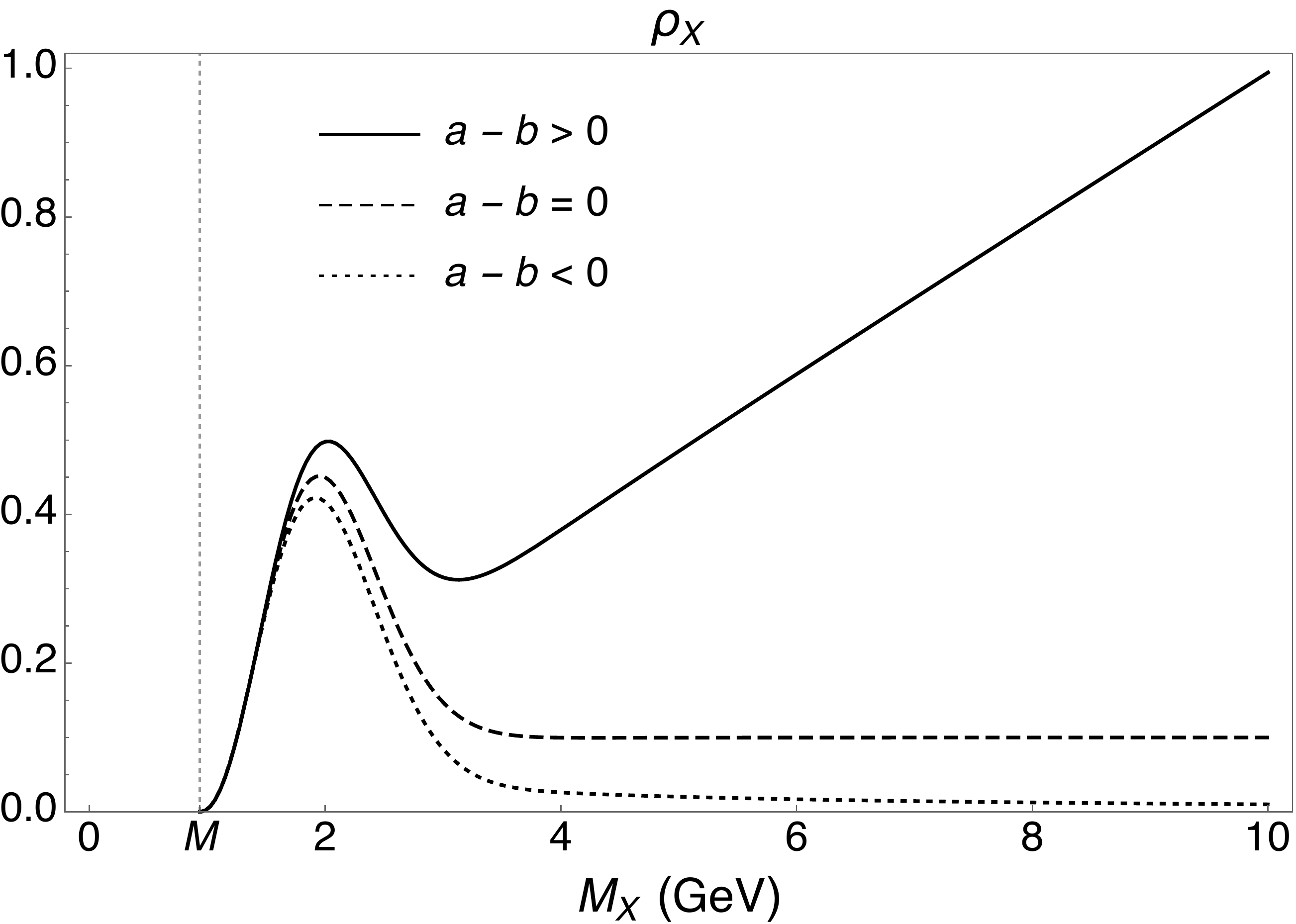}
 \caption{The spectral function $\rho_X$ of Eq.~\eqref{eq:spectral} as a function of the spectator mass $M_X$.}
\label{fig:spectral}
\end{figure}

The gluon TMDs of Eqs.~\eqref{eq:f1}-\eqref{eq:h1perp} explicitly depend on the spectator mass $M_X$, which therefore must not be considered as a free parameter. In fact, in our model $M_X$ can take real values in a continuous range according to the spectral function
\begin{equation}
 \rho_X (M_X) =  \mu^{2a} \left[ \frac{A}{B + \mu^{2b}} + \frac{C}{\pi \sigma} e^{- \frac{(M_X - D)^2}{\sigma^2}} \right] \; ,
 \label{eq:spectral}
\end{equation}
where $\mu^2 = M_X^2 - M^2$ and $\{ X \} \equiv \{ A, B, a, b, C, D, \sigma \}$ are free parameters. 
Indeed, each gluon TMD in Eqs.~\eqref{eq:f1}-\eqref{eq:h1perp} is weighed on the spectral function $\rho_X$ such that the actual model expression of a generic gluon TMD $F^g (x, \bm{p}_T^2)$ reads 
\begin{equation}
F^g (x, \bm{p}_T^2) = \int_M^\infty dM_X \, \rho_X ( M_X ) \, \hat{F}^g (x, \bm{p}_T^2; M_X) \; .
\label{eq:weightedTMD}	
\end{equation}
As shown in Fig.~\ref{fig:spectral}, the spectral function is particularly sensitive to the parameters $a, b$: its asymptotic trend at large $M_X$ depends on the sign of the difference $a-b$. As pointed out in Ref.~\cite{Goldstein:2010gu}, it is easy to show that the trend at large $M_X$ affects the small-$x$ tail of TMDs, which is the effective way in our model to account for $q\bar{q}$ contributions to spectator configurations that become energetically available at large $M_X$. Similarly, the behavior of $\rho_X$ at low $M_X$ influences the tail of TMDs at intermediate $x$.

\section{Model parameters}
\label{s:params}

According to Eq.~\eqref{eq:weightedTMD}, our model results for the $T$-even gluon TMDs are obtained by weighing the analytic expressions of
Eqs.~\eqref{eq:f1}-\eqref{eq:h1perp} with the spectral function in Eq.~\eqref{eq:spectral}.
In total, we have 10 free parameters: seven characterizing the spectral function ($A$, $B$, $a$, $b$, $C$, $D$, $\sigma$) and three for the dipolar
form factor (the $\kappa_{1,2}$ normalizations and the $\Lambda_X$ cut-off). 
The parameters $A$, $B$ and $b$, in the Lorentzian component of the spectral function~\eqref{eq:spectral} control the small-$x$ tail of the gluon TMDs. The Gaussian component (depending on the parameters $C$, $D$ and $\sigma$) is sensitive to the moderate-$x$ regime. The sum of the two contributions is modulated by a power-law behavior (depending on the parameter $a$) such that the spectral function has enough flexibility to correctly describe the whole $x$-range considered. The vertex parameters $\kappa_{1,2}$ and $\Lambda_X$ in Eq.~\eqref{eq:dipolarff} mainly regulate the behavior in $\bm{p}_T$.
To fix these parameters, we follow the procedure described below. 

We perform the integration over $\bm{p}_T$ in the TMDs of Eqs.~\eqref{eq:f1}-\eqref{eq:h1perp}
weighed with the spectral function as in Eq.~\eqref{eq:weightedTMD}.
As is well known, only the first two densities give non-vanishing
results. Then, we assume that these $\bm{p}_T$-integrated TMDs reproduce the
collinear PDFs $f_1^g (x)$ and $g_1^g (x)$ at some low scale $Q_0$.
Finally, we fix our model parameters by simultaneously fitting the {\tt NNPDF3.1sx} parametrization for $f_1^g$~\cite{Ball:2017otu} and the {\tt NNPDFpol1.1} parametrization for $g_1^g$~\cite{Nocera:2014gqa} at $Q_0 = 1.64$ GeV, which is the lowest hadronic scale provided by the {\tt NNPDF} Collaboration for these two parametrizations. 
We consider the parametrizations only in the range
$0.001 < x < 0.7$ to avoid regions with large uncertainties and where effects
not included in our model can be relevant~\cite{Ball:2014uwa}. We choose a
grid of 70 points distributed logarithmically below $x=0.1$ and 30 points
distributed linearly above $x=0.1$.  
We perform the fit using the bootstrap method. Namely, we create $N$ replicas
of the central value of the {\tt NNPDF} parametrization by randomly altering
it with a Gaussian noise with the same variance as the original
parametrization uncertainty. We then fit each replica separately and we obtain
a vector of $N$ results for each model parameter.
We build the 68\% uncertainty band of our fit by rejecting
the largest and smallest 16\% of the $N$ values of any prediction.
This 68\% band corresponds to the $1\sigma$ confidence level only if the
predictions follow a Gaussian distribution, which is not true in general.
We choose to work with $N=100$ replicas because this number is
sufficient to reproduce the uncertainty of the original {\tt NNPDF}
parametrization.
In the following, we will show also the result from the replica number 11,
which we consider a particularly representative replica because its parameter
values are the closest to the mean parameters. However, we stress that only the
full set of 100 replicas contain the full information about our fit
results.~\footnote{The full set of results can be obtained from the authors
  upon request.}

 \begin{table}
 \begin{tabular}[c]{c|c|c}
 \hline \hline
 parameter                   &    mean                         & replica 11 \\
 \hline			
 $A$                             & 6.1 $\pm$ 2.3       & 6.0            \\
 $a$                              & 0.82 $\pm$ 0.21       & 0.78        \\
 $b$                              & 1.43 $\pm$ 0.23       & 1.38        \\
 $C$                             & 371 $\pm$ 58 & 346     \\
 $D$ (GeV)                   & 0.548 $\pm$ 0.081     & 0.548         \\
 $\sigma$ (GeV)           & 0.52 $\pm$ 0.14       & 0.50         \\
 $\Lambda_X$ (GeV)   & 0.472 $\pm$ 0.058       & 0.448         \\
 $\kappa_1$ (GeV$^2$)       & 1.51 $\pm$ 0.16       & 1.46         \\
 $\kappa_2$ (GeV$^2$)       & 0.414 $\pm$ 0.036       & 0.414         \\
 \hline \hline
  \end{tabular}
 \caption{Central column: mean values and uncertainties of the fitted model parameters. Rightmost column: corresponding values for replica 11.}
 \label{tab:par-g}
 \end{table}

In Tab.~\ref{tab:par-g}, we show the values of our model parameters. For each one, we quote the central 68\% of the $N=100$ values by indicating the average
and the uncertainty given by the semi-difference of the upper and lower limits. In the right column, we show the corresponding values for replica 11.
Parameter $B$ in Eq.~\eqref{eq:spectral} is fixed to $B = 2.1$ since
exploratory tests have shown that the fit is rather insensitive to it.  We get
a total $\chi^2$/d.o.f. = 0.54 $\pm$ 0.38. This small value
originates from the large uncertainty in the $g_1^g$ parametrization, particularly at small $x$. 
We remark that the output of the fit selects the option $a-b < 0$, which corresponds to a spectral function 
$0 \leq \rho (M_X) < 0.5$ asymptotically vanishing for very large spectator masses $M_X$ (see Fig.~\ref{fig:spectral}). Therefore, we deduce that the positivity bound fulfilled by $\hat{F}^g$ in the right handside of Eq.~\eqref{eq:weightedTMD} (when corresponding to the polarized TMDs of Eqs.~\eqref{eq:g1L}-\eqref{eq:h1perp}) is maintained through the integral also for the actual gluon TMDs on the left handside.

\begin{figure}[h]
\includegraphics[width=0.42\textwidth]{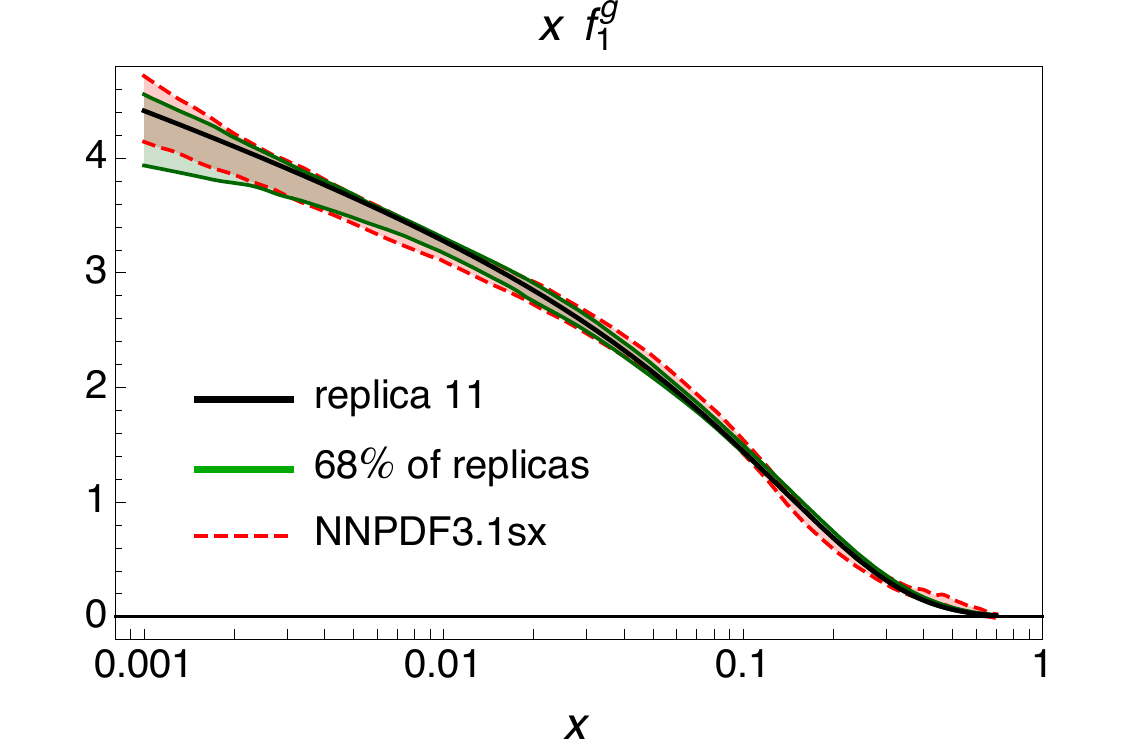}
\hspace{0.4cm}
\includegraphics[width=0.42\textwidth]{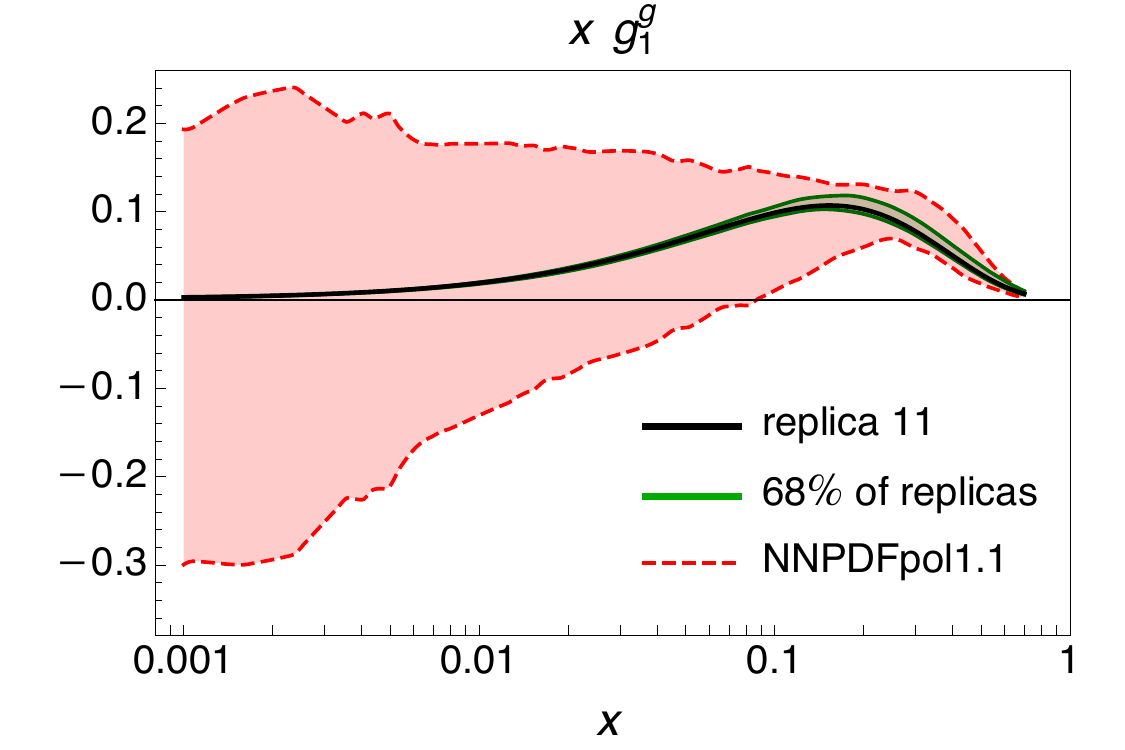}
\caption{The $x f_1^g$ (left panel) and $x g_1^g$ (right panel) as functions of $x$ at $Q_0 = 1.64$ GeV. Lighter band with red dashed borders for the {\tt NNPDF3.1sx} parametrization of $x f_1^g$~\cite{Ball:2017otu} and the {\tt NNPDFpol1.1} parametrization of $x g_1^g$~\cite{Nocera:2014gqa}. Green band for the 68\% uncertainty band of the spectator model fit. Solid black line for the result of the replica 11.}
\label{fig:PDF-g}
\end{figure}

In Fig.~\ref{fig:PDF-g}, we show the results of our simultaneous fit of $x
f_1^g (x)$ (left panel) and $x g_1^g (x)$ (right panel) at $Q_0 = 1.64$
GeV. The lighter band with red dashed borders identifies the {\tt
  NNPDF3.1sx} parametrization of $x f_1^g$~\cite{Ball:2017otu} and the {\tt
  NNPDFpol1.1} parametrization of $x g_1^g$~\cite{Nocera:2014gqa}. The green
band is the 68\% uncertainty band of our fit. The solid black line represents
the result of replica 11.
The right panel shows that our gluon helicity at most diverges more slowly
than $1/x$. On the one side, this feature can ben considered as a rigidity of
the model. On the other side, it can be considered as a prediction.
In any case, we verified that it is important to perform a simultaneous fit of
both the unpolarized and helicity gluon PDFs. Bounding the model parameters
only to $f_1^g (x)$ is not enough to get a reliable $x$-behavior of the model.

\section{Results}
\label{s:out}

With the parameters in Tab.~\ref{tab:par-g}, the second Mellin moment of our model PDF $f_1^g (x, Q_0)$, \emph{i.e.}, the nucleon momentum fraction carried by the gluons at the model scale $Q_0 = 1.64$ GeV, turns out to be
\begin{equation}
\langle x \rangle_g = \int_{0}^1 dx \, x \, f_1^g (x, Q_0) = 0.424 \pm 0.009 \; .
\label{eq:x_avg}
\end{equation}
This result is in excellent agreement with the latest lattice calculation $\langle x \rangle_g = 0.427(92)$ obtained at the scale 2 GeV~\cite{Alexandrou:2020sml}.
The first Mellin moment of the model PDF $g_1^g (x)$ gives the contribution of the gluon helicity to the nucleon spin. In our model, it turns out to be
$S_g = \textstyle{\frac{1}{2}} \langle 1 \rangle_{\Delta g} = 0.159 \pm 0.011$ at $Q_0 = 1.64$ GeV, to be compared with the latest lattice estimate of the gluon total angular momentum $\langle J \rangle_g = 0.187(46)$ at the scale 2 GeV~\cite{Alexandrou:2020sml}. 

\begin{figure}[h]
\includegraphics[width=0.8\textwidth]{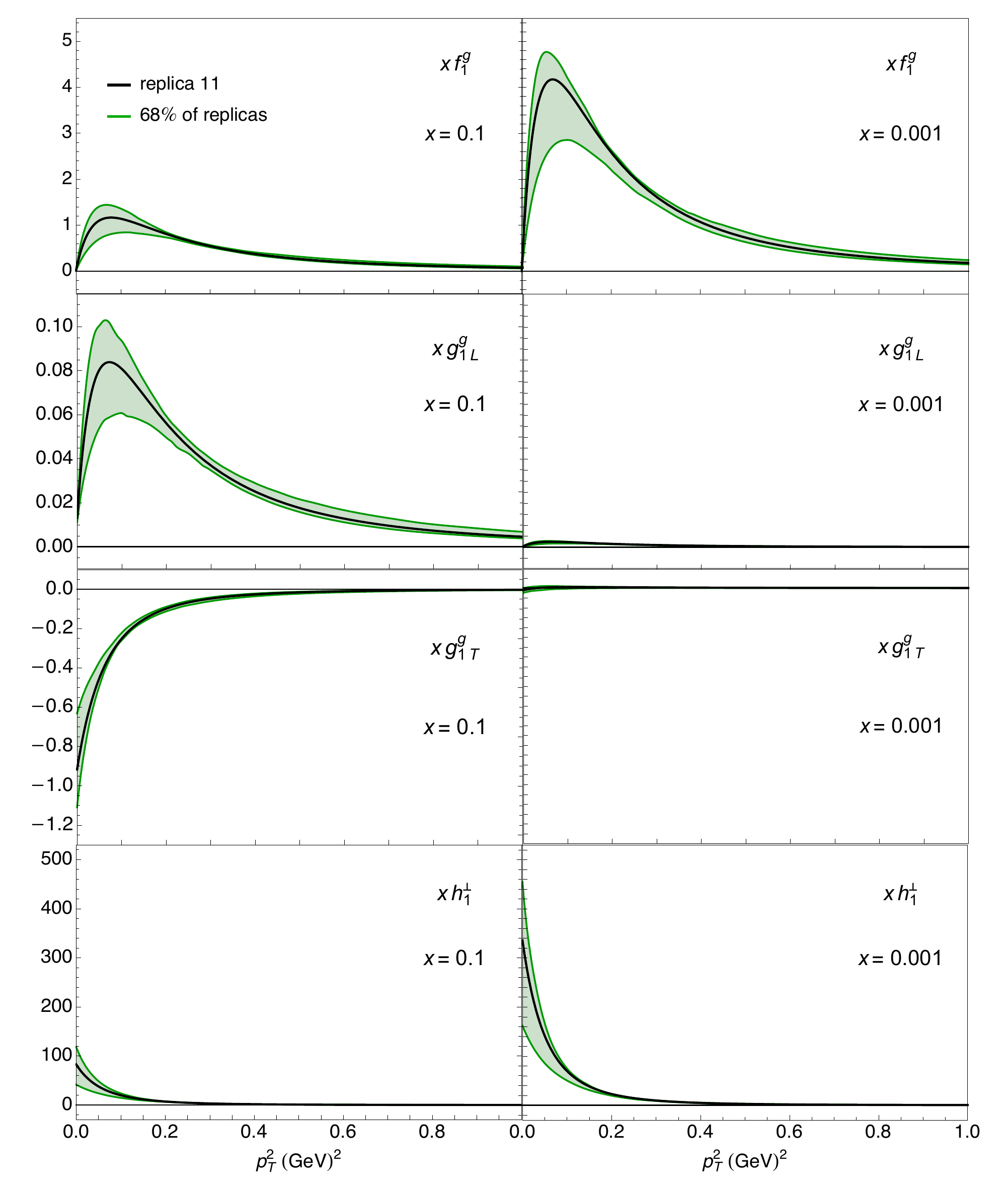} 
\caption{The $T$-even gluon TMDs as functions of $\bm{p}_T^2$ for $x=0.1$ (left panels) and $x=0.001$ (right panels) at $Q_0 = 1.64$ GeV. Green band indicates the 68\% statistical uncertainty, solid black line for the replica 11. From top to bottom, panels show $x f_1^g (x, \bm{p}_T^2)$, $x g_{1L}^g (x, \bm{p}_T^2)$, $x g_{1T}^g (x, \bm{p}_T^2)$, and $x h_1^{\perp g} (x, \bm{p}_T^2)$.}
\label{fig:TMD-g}
\end{figure}

In Fig.~\ref{fig:TMD-g}, we show our model results for $T$-even gluon TMDs as functions of $\bm{p}_T^2$ for $x=0.1$ (left panels) and $x=0.001$ (right panels) at the same scale $Q_0 = 1.64$ GeV as in Fig.~\ref{fig:PDF-g}, \emph{i.e.}, without evolution effects. Again, the green band refers to the 68\% statistical uncertainty, and the solid black line indicates the result of the best replica 11. From top to bottom, the panels refer to the unpolarized $x f_1^g (x, \bm{p}_T^2)$, the helicity $x g_{1L}^g (x, \bm{p}_T^2)$, the worm-gear $x g_{1T}^g (x, \bm{p}_T^2)$, and the Boer--Mulders $x h_1^{\perp g} (x, \bm{p}_T^2)$. Each TMD shows a distinct pattern both in $x$ and $\bm{p}_T^2$. In particular, the unpolarized $x f_1^g (x, \bm{p}_T^2)$ clearly shows a non-Gaussian shape in $\bm{p}_T^2$ with a large flattening tail for $\bm{p}_T^2 \to 1$ GeV. Moreover, for $\bm{p}_T^2 \to 0$ it reaches a very small but non-vanishing value, suggesting that the gluon wave function has a significant component with orbital angular momentum $L=1$~\footnote{This result would change if the spectator were a particle with spin different from $\textstyle{\frac{1}{2}}$.}. The information underlying these plots largely expands the one contained in Fig.~\ref{fig:PDF-g} and can be a useful guidance in explorations of the full 3D dynamics of gluons.

To this purpose, it is also useful to consider the following densities that describe the 2D $\bm{p}_T$-distribution of gluons at different $x$ for various combinations of their polarization and of the nucleon spin state. For an unpolarized nucleon, we identify the unpolarized density 
\begin{equation}
 x \rho (x, p_x, p_y) = x f_1^g (x, \bm{p}_T^2) 
\label{eq:rho_unpol}
\end{equation}
as the probability density of finding unpolarized gluons at given $x$ and $\bm{p}_T$, while the ``Boer--Mulders" density 
\begin{equation}
 x \rho^{\leftrightarrow} (x, p_x, p_y) = \frac{1}{2} \bigg[ x f_1^g (x, \bm{p}_T^2) + \frac{p_x^2 - p_y^2}{2 M^2} \, x h_1^{\perp g} (x, \bm{p}_T^2) \bigg]
\label{eq:rho_T}
\end{equation}
represents the probability density of finding gluons linearly polarized in the transverse plane at $x$ and $\bm{p}_T$. The ``helicity density"  
\begin{equation}
 x \rho^{\circlearrowleft/+} (x, p_x, p_y) =  x f_1^g (x, \bm{p}_T^2) + x g_{1L}^g (x, \bm{p}_T^2) 
 \label{eq:rho_circle_+}
\end{equation}
contains the probability density of finding circularly polarized gluons at $x$ and $\bm{p}_T$ in longitudinally polarized nucleons. Finally, the ``worm-gear density"
\begin{equation}
 x \rho^{\circlearrowleft/\rightarrow} (x, p_x, p_y) =  x f_1^g (x, \bm{p}_T^2) - \frac{p_x}{M} \, x g_{1T}^g (x, \bm{p}_T^2) 
\label{eq:rho_circle_T}
\end{equation}
is similar to the previous one but for transversely polarized nucleons. The first and third densities describe a situation where the $\bm{p}_T$-distribution is cylindrically symmetric around the longitudinal direction identified by $P^+$, because the nucleon (gluon) is unpolarized or polarized longitudinally (circularly) along $P^+$. The density in Eq.~\eqref{eq:rho_T} is symmetric about the $p_x$ and $p_y$ axes because it describes unpolarized nucleons and gluons that are linearly polarized along the $p_x$ direction. The density in Eq.~\eqref{eq:rho_circle_T} involves a transverse polarization of the nucleon along the $+p_x$ axis. Hence, we expect it to display an asymmetric distribution in the same direction.

\begin{figure}[h]
\includegraphics[width=0.8\textwidth]{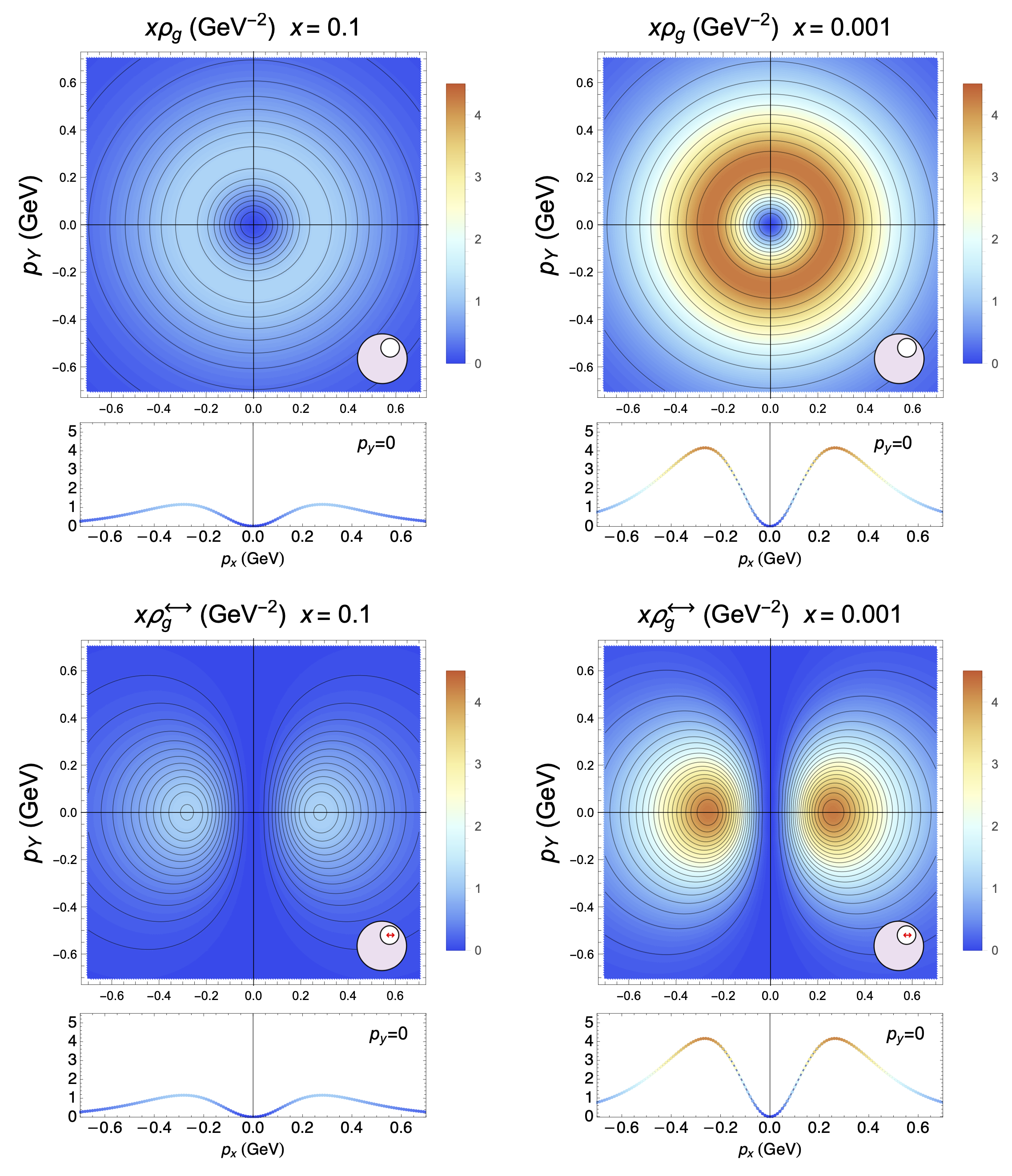} 
 \caption{From top to bottom, the gluon densities of Eqs.~\eqref{eq:rho_unpol} and~\eqref{eq:rho_T} as functions of $\bm{p}_T$ at $Q_0 = 1.64$ GeV and at $x=0.1$ (left panels) and $x=0.001$ (right panels) for an unpolarized nucleon virtually moving towards the reader. For each contour plot, 1D ancillary plots show the density at $p_y = 0$. Results from the best replica 11 (see text).}
\label{fig:contour-Nunpol}
\end{figure}

In Fig.~\ref{fig:contour-Nunpol}, from top to bottom the contour plots show the $\bm{p}_T$-distribution of the densities in Eqs.~\eqref{eq:rho_unpol} and~\eqref{eq:rho_T}, respectively, obtained at $Q_0 = 1.64$ GeV from replica 11 at $x=0.1$ (left panels) and $x=0.001$ (right panels) for an unpolarized nucleon virtually moving towards the reader. The color code identifies the size of the oscillation of each density along the $p_x$ and $p_y$ directions. In order to better visualize these oscillations, ancillary 1D plots are shown below each contour plot, which represent the corresponding density at $p_y = 0$. As expected, the density of Eq.~\eqref{eq:rho_unpol} (top panels) has a cylindrical symmetry around the direction of motion of the nucleon pointing towards the reader. Since the nucleon is unpolarized but the gluons are linearly polarized along the $p_x$ direction, the density of Eq.~\eqref{eq:rho_T} (bottom-row panels) shows a quadrupole structure. This departure from the cylindrical symmetry is emphasized at small $x$, because the Boer--Mulders function is particularly large.

\begin{figure}[h]
\includegraphics[width=0.8\textwidth]{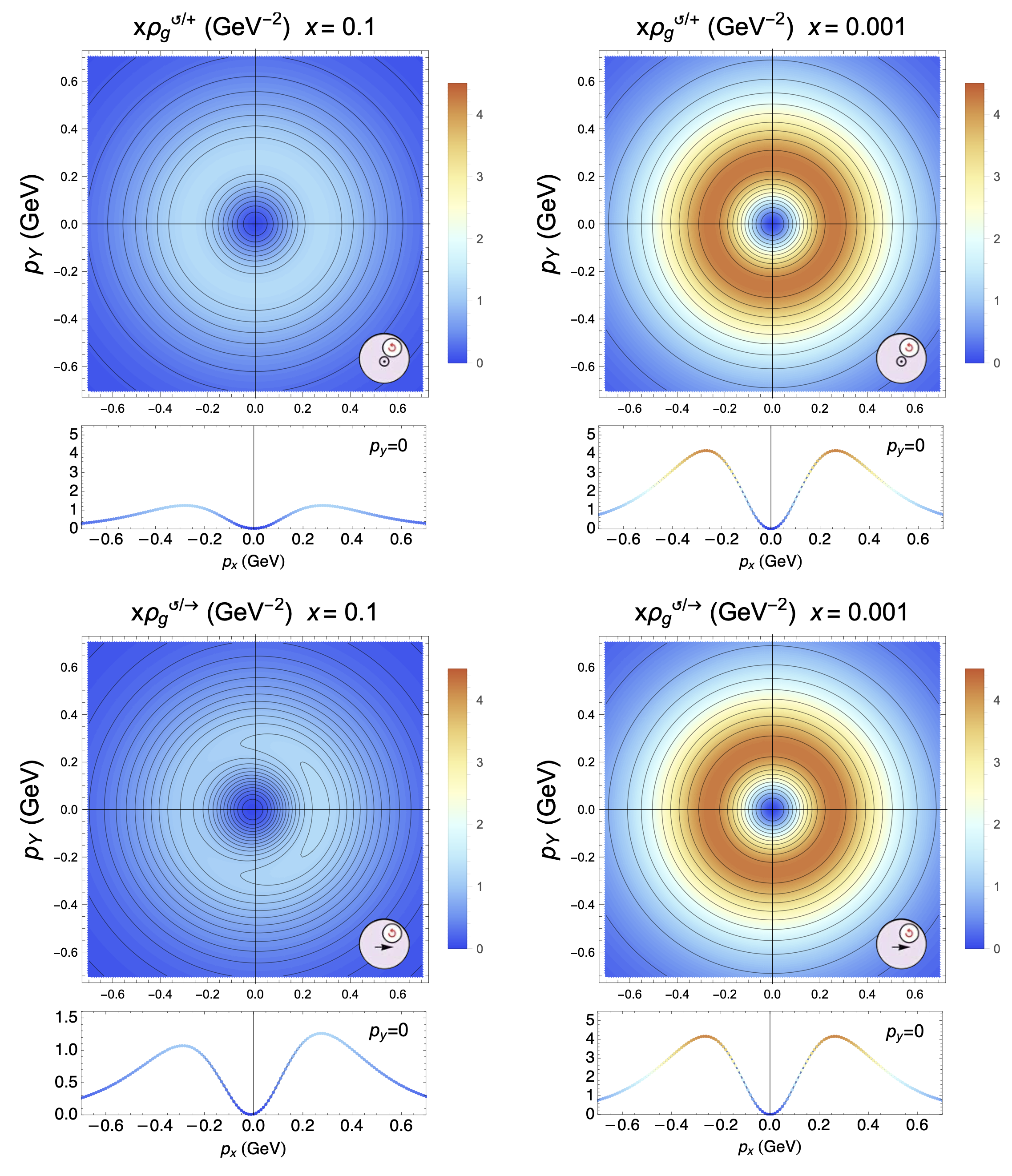}
\caption{From top to bottom, the gluon densities of Eqs.~\eqref{eq:rho_circle_+} and ~\eqref{eq:rho_circle_T} as functions of $\bm{p}_T$ at $Q_0 = 1.64$ GeV and at $x=0.1$ (left panels) and $x=0.001$ (right panels) for a polarized nucleon virtually moving towards the reader. For each contour plot, 1D ancillary plots show the density at $p_y = 0$. Results from the best replica 11 (see text).}
\label{fig:contour-Npol}
\end{figure}

In Fig.~\ref{fig:contour-Npol}, from top to bottom the plots show the $\bm{p}_T$-distribution of the densities in Eqs.~\eqref{eq:rho_circle_+}
and~\eqref{eq:rho_circle_T}, respectively, obtained at $Q_0 = 1.64$ GeV from replica 11 at $x=0.1$ (left panels) and $x=0.001$ (right panels) for a polarized nucleon virtually moving towards the reader. Color code and notations are the same as in the previous figure. The $x \rho^{\circlearrowleft/+}$ density of Eq.~\eqref{eq:rho_circle_+} (top panels) is perfectly symmetric in the displayed transverse plane because it refers to a nucleon (gluon) longitudinally (circularly) polarized along the direction of motion pointing towards the reader. The size of the density is emphasized at smaller $x$. The $x \rho^{\circlearrowleft/\rightarrow}$ density of
Eq.~\eqref{eq:rho_circle_T} is slightly asymmetric in $p_x$ at $x=0.1$ (left bottom panel) because the nucleon is transversely polarized along the $p_x$
direction. This asymmetry is small and vanishes at $x=0.001$ (right bottom panel) because  of the behavior of the worm-gear function $g_{1T}$.

\section{Conclusions and Outlook}
\label{s:end}

We presented a systematic calculation of leading-twist $T$-even gluon TMDs under the assumption that what remains of a nucleon after emitting a gluon can be  effectively treated as a single spin-$\textstyle{\frac{1}{2}}$ spectator particle. The latter is considered on-shell but its mass is allowed to take a continuous range of values described by a spectral function. The model parameters are fixed by reproducing the $x$-profile of collinear unpolarized and helicity gluon PDFs extracted from global fits. Nevertheless, the spectral function grants the model a sufficient degree of flexibility and gives the opportunity of actually incorporating the effect of $q\bar{q}$ contributions, which are normally absent in spectator models.

We discussed our model results for the tomography in momentum space of gluons
inside nucleons for various combinations of their polarizations. These results
can be a useful guidance to the investigation of observables sensitive to
gluon TMD dynamics, with applications ranging from
heavy-flavor-meson/open-charm/quarkonia/Higgs production (see, {\it e.g.},
Refs.~\cite{Boer:2011kf,Boer:2016fqd,Godbole:2014tha,Mukherjee:2016qxa,Bacchetta:2018ivt,Godbole:2017syo,DAlesio:2017rzj,DAlesio:2019gnu,Gutierrez-Reyes:2019rug,Dunnen:2014eta})
to almost back-to-back di-hadron and di-jet production (see, {\it e.g.},
Refs.~\cite{Boer:2016fqd,Zheng:2018awe}), and almost back-to-back $J/\psi$-jet
production~\cite{DAlesio:2019qpk}. All these channels can be studied at
current and future collider facilities. In order to facilitate the computation
of observables based on our model, we will make our results available
through the {\tt TMDlib} library~\cite{Hautmann:2014kza}.

We plan to extend our model to include leading-twist $T$-odd gluon TMDs along lines similar to our previous work on quark TMDs~\cite{Bacchetta:2008af}. It would be interesting also to improve the description of the unpolarized gluon TMD at small $x$ such that it simultaneously satisfies evolution equations in both the Collins--Soper--Sterman (CSS)~\cite{Collins:1981uk,Collins:2011zzd} and BFKL~\cite{Fadin:1975cb,Kuraev:1976ge,Kuraev:1977fs,Balitsky:1978ic} kinematical regimes. All these prospective developments are relevant to the exploration of the gluon dynamics inside nucleons and nuclei, which constitutes one of the major goals of the \emph{Electron-Ion Collider} (EIC) project~\cite{Boer:2011fh,Accardi:2012qut}.

\begin{acknowledgments}
This work is supported by the European Research Council (ERC) under the European Union's Horizon 2020 research and innovation program (grant agreement
No. 647981, 3DSPIN) and by the Italian MIUR under the FARE program (code n. R16XKPHL3N, 3DGLUE). 
\end{acknowledgments}




\bibliographystyle{apsrevM}
\bibliography{gluonspectator_revised}


\end{document}